\documentclass[a4paper,10pt]{article}
\usepackage{fullpage,epsfig}

\begin{document}

\title{The Kondo effect in periodic narrow-band systems}
\author{V.Yu.~Irkhin$^{*}$ and A.V.~Zarubin \\
Institute of Metal Physics, 620219 Ekaterinburg, Russia}
\maketitle

\begin{abstract}
The Kondo divergences owing to interaction of current carriers with local
moments in highly correlated electron systems are considered within the
Hubbard and $s$-$d$ exchange models with infinitely strong on-site
interaction, the many-electron Hubbard representation being used. The picture
of density of states containing a peak at the Fermi level is obtained.
Various forms of the self-consistent approximation are used. The problem of
the violation of analytical properties of the Green's function is discussed.
Smearing of the ``Kondo'' peak owing to spin dynamics and finite temperatures
is investigated.

\end{abstract}

The problem of strong correlations and magnetism in many-electron systems is
one of the most important in the solid state theory. Since the Hubbard's
works of 60's, a great progress has been achieved in understanding
electronic structure of systems with strong on-site interaction. Last time,
the role of the Kondo effect has been discussed within the large-$d$
approach ($d$ is space dimensionality) which reduces the initial periodic
Hubbard model to an effective Anderson impurity model
\cite{large-d,Georges:1996}. Besides the Hubbard bands, an important role in the
formation of density of states (DOS) picture belongs to a peak at the Fermi
level, which was found for both half-filled and doped case (the latter case
is considered in \cite{Georges:1996,Anisimov:1997}). It should be noted that
this approach meets with a number of computational difficulties (e.g.,
consideration of finite temperatures is needed, and the low-temperature
limit is non-trivial). The structure of the spectrum in large-$d$ approaches
is confirmed by the Monte-Carlo calculations. On the other hand, this
feature was not reproduced by most preceding analytical approaches. In
particular, the Hubbard-III approximation \cite{Hubbard-III:1964} does not
take into account contributions of Fermi-like excitations in a proper way
because of its single-site character. A detailed analysis of this
approximation was performed in Refs.~\cite{Anokhin:1991,Anokhin:1991a}
within the large-$z$ ($z$ is nearest-neighbor number) expansion.

In the present paper we present a treatment that is based on the method of
equations of motion for the many-electron Hubbard operators
\cite{Hubbard-IV:1965,Irkhin:1994} and is much more simple than the large-$d$
approach. As the zero order this approach reduces to the simplest Hubbard-I
approximation. General expressions for $1/z$-corrections were obtained in
Ref.~\cite{Anokhin:1991}. Unfortunately, the terms with the one-particle
occupation numbers (which just describe the Kondo effect) were neglected in
Refs.~\cite{Anokhin:1991,Anokhin:1991a}, and only a classical approximation
was considered in Ref.~\cite{Anokhin:1991a}.

We start from the $s$-$d$ exchange model with the large $s$-$d$ coupling
parameter $|I|$,
\begin{equation}
\mathcal{H}=\sum_{\mathbf{k}\sigma }t_{\mathbf{k}}c_{\mathbf{k}\sigma
}^{\dagger }c_{\mathbf{k}\sigma }-I\sum_{i\sigma \sigma ^{\prime }}\mathbf{S}
_i\mbox {\boldmath $\sigma $}_{\sigma \sigma ^{\prime }}c_{i\sigma
}^{\dagger }c_{i\sigma ^{\prime }}+\mathcal{H}_d,  \label{eq:H0}
\end{equation}
where $t_{\mathbf{k}}$ is the band energy, $\mathcal{H}_d$ is the Heisenberg
Hamiltonian of the localized-spin system, {\boldmath $\sigma $} are the
Pauli matrices. In the limit $|I|\rightarrow \alpha \infty $ (here and
hereafter $\alpha =\mathop{\mathrm{sign}}I=\pm $), it is convenient to pass
to the atomic representation of the Hubbard operators $X_i^{\beta \gamma
}=|i\beta \rangle \langle i\gamma |$ where $\mathcal{H}_{sd}$ (the second
term in (\ref{eq:H0})) takes the diagonal form \cite{sdem}. For the electron
concentration $n<1,$ after performing the procedure of projection onto the
corresponding state space, the one-electron Fermi operators $c_{i\sigma
}^{\dagger }$ are replaced by the many-electron operators $g_{i\sigma \alpha
}^{\dagger }.$ These are expressed in terms of the $X$-operators as
\begin{eqnarray*}
g_{i\sigma +}^{\dagger } &=&\sum_M\{(S+\sigma M+1)/(2S+1)\}^{1/2}X_i(M+\frac
\sigma 2,+;M), \\
g_{i\sigma -}^{\dagger } &=&\sum_M\sigma \{(S-\sigma
M)/(2S+1)\}^{1/2}X_i(M+\frac \sigma 2,-;M),
\end{eqnarray*}
where $|M\rangle $ are the empty states and $|m\alpha \rangle $ are the
singly-occupied states with the total on-site spin $S+\alpha /2$ and its
projection $m$ (which survive at $|I|\rightarrow \alpha \infty $), $\sigma
=\pm $. The Hamiltonian of the $s$-$d$ exchange interaction yields a
constant energy shift only and can be omitted, so that we obtain
\begin{equation}
\mathcal{H}=\sum_{\mathbf{k}\sigma }t_{\mathbf{k}}g_{\mathbf{k}\sigma \alpha
}^{\dagger }g_{\mathbf{k}\sigma \alpha }+\mathcal{H}_d,\qquad \alpha =
\mathop{\mathrm{sign}}I.  \label{eq:H}
\end{equation}
For $n>1$ we have to pass to the ``hole'' representation by introducing new
localized spins $\widetilde{S}=S\pm 1/2$, and the Hamiltonian takes the same
form (\ref{eq:H}) with the replacement $t_{\mathbf{k}}\rightarrow
-t_{\mathbf{k}}(2\widetilde{S}+1)/(2S+1)$. For $S=1/2$, $I\rightarrow -\infty $
the Hamiltonian (\ref{eq:H}) coincides with that of the Hubbard model in the
case of the infinite on-site repulsion $U\rightarrow \infty $ with the
replacement $t_{\mathbf{k}}\rightarrow t_{\mathbf{k}}/2$ and the almost
half-filled band, so that $n\rightarrow \delta $ with $\delta $ the hole
concentration. Thus we do not need to discuss the Hubbard model separately
(physically, both models describe the local-moment situation).

We calculate the one-particle Green's function for ferromagnetic (FM) state
in the energy representation,
\begin{equation}
G_{\mathbf{k}\sigma \alpha }(E)=\langle \!\langle g_{\mathbf{k}\sigma \alpha
}|g_{\mathbf{k}\sigma \alpha }^{\dagger }\rangle \!\rangle _E.
\label{eq:EGF:0}
\end{equation}
Using the commutation relations for for the operators $g_{i\sigma \alpha }$
and spin operators (cf. Refs.~\cite{Anokhin:1991,Anokhin:1991a}) we obtain
the equation of motion
\begin{eqnarray}
(E-t_{\mathbf{k-q}\sigma \alpha })G_{\mathbf{k}\sigma \alpha }(E)
&=&P_{\sigma \alpha }+\frac \alpha {2S+1}\sum_{\mathbf{q}}t_{\mathbf{k-q}}
\{\langle \!\langle \delta [\sigma S_{\mathrm{tot}\mathbf{q}}^z-\frac
12\sum_MX_{\mathbf{q}}(M\alpha ;M\alpha )]g_{\mathbf{k-q}\sigma \alpha }|
g_{\mathbf{k}\sigma \alpha }^{\dagger }\rangle \!\rangle _E  \nonumber \\
&&+\langle \!\langle S_{\mathrm{tot}\mathbf{q}}^{-\sigma }g_{\mathbf{k-q},
-\sigma \alpha }|g_{\mathbf{k}\sigma \alpha }^{\dagger }\rangle \!\rangle
_E\},  \label{eq:eqm1}
\end{eqnarray}
with $S_{\mathrm{tot}}^{\pm }$, $S_{\mathrm{tot}}^z$ being the total spin
operators (including the contributions of empty and singly-occupied states,
see for details \cite{Anokhin:1991a}),
\[
t_{\mathbf{k}\sigma \alpha }=P_{\sigma \alpha }t_{\mathbf{k}},\qquad
P_{\sigma \alpha }=\frac{\widetilde{S}+1/2+\alpha \sigma \langle
S_{\mathrm{tot}}^z\rangle -\alpha n/2}{2S+1},
\]
$\langle S_{\mathrm{tot}}^z\rangle $ is the total on-site average
magnetization. The ``Kondo'' term comes from the last Green's function in
the right-hand side of (\ref{eq:eqm1}), which describes spin-flip processes.
Performing decoupling of the next equation of motion to first order in the
nearest-neighbor number $1/z$ we derive
\begin{eqnarray}
(E-t_{\mathbf{k-q}\sigma \alpha })\langle \!\langle S_{\mathrm{tot}\mathbf{q}%
}^{-\sigma }g_{\mathbf{k-q},-\sigma \alpha }|g_{\mathbf{k}\sigma \alpha
}^{\dagger }\rangle \!\rangle _E &=&\frac \alpha {2S+1}[\chi _{\mathbf{q}%
\sigma }+(2S+1)n_{\mathbf{k-q}-\sigma \alpha }](1+t_{\mathbf{k}}G_{\mathbf{k}%
\sigma \alpha }(E))  \nonumber \\
&&-\alpha n_{\mathbf{k-q}-\sigma \alpha }t_{\mathbf{k-q}}G_{\mathbf{k}\sigma
\alpha }(E).  \label{eq:eqm2}
\end{eqnarray}
Here we have neglected $\mathcal{H}_d$,
\[
\chi _{\mathbf{q}\sigma }=\langle S_{\mathrm{tot}\mathbf{-q}}^\sigma S_{%
\mathrm{tot}\mathbf{q}}^{-\sigma }\rangle ,\qquad n_{\mathbf{k}\sigma \alpha
}=\langle g_{\mathbf{k}\sigma \alpha }^{\dagger }g_{\mathbf{k}\sigma \alpha
}\rangle .
\]
To lowest order (i.e. in the Hubbard-I approximation) we have
\[
n_{\mathbf{k}\sigma \alpha }=P_{\sigma \alpha }f(t_{\mathbf{k}\sigma \alpha
})
\]
where $f(E)$ is the Fermi function. After substituting (\ref{eq:eqm2}) and a
similar equation for the longitudinal contribution in (\ref{eq:eqm1}) (which
contains $\chi _{\mathbf{q}}^{zz}=\langle S_{\mathrm{tot}\mathbf{-q}}^zS_{%
\mathrm{tot}\mathbf{q}}^z\rangle $) into (\ref{eq:eqm1}) we obtain
\begin{equation}
G_{\mathbf{k}\sigma \alpha }(E)=\frac{a_{\mathbf{k}\sigma \alpha }(E)}{b_{%
\mathbf{k}\sigma \alpha }(E)-a_{\mathbf{k}\sigma \alpha }(E)t_{\mathbf{k}}},
\label{eq:EGF}
\end{equation}
where
\begin{eqnarray}
a_{\mathbf{k}\sigma \alpha }(E) &=&P_{\sigma \alpha }+\sum_{\mathbf{q}}\frac{%
t_{\mathbf{k-q}}}{(2S+1)^2}\left[ \frac{\chi _{\mathbf{q}\sigma }+(2S+1)n_{%
\mathbf{k-q}-\sigma \alpha }}{E-t_{\mathbf{k-q}-\sigma \alpha }}+\frac{\chi
_{\mathbf{q}}^{zz}}{E-t_{\mathbf{k-q}\sigma \alpha }}\right] ,  \label{eq:a}
\\
b_{\mathbf{k}\sigma \alpha }(E) &=&E-\sum_{\mathbf{q}}\frac{t_{\mathbf{k-q}%
}^2}{2S+1}\frac{n_{\mathbf{k-q}-\sigma \alpha }}{E-t_{\mathbf{k-q}-\sigma
\alpha }}.  \label{eq:b}
\end{eqnarray}
The longitudinal fluctuations can be neglected in the saturated FM region
\cite{Zarubin:1999}, but are important in the paramagnetic region (the
corresponding results are obtained from the above formulas at $\langle S_{%
\mathrm{tot}}^z\rangle =0$) which is considered hereafter. The equation for
the chemical potential reads
\[
\sum_{\mathbf{k}}\langle g_{\mathbf{k}\sigma \alpha }^{\dagger }g_{\mathbf{k}%
\sigma \alpha }\rangle =\frac n2.
\]

To smear the logarithmic DOS singularity at the Fermi level, one has to take
into account spin dynamics which is determined by the Hamiltonian $\mathcal{H%
}_d$ (cf.~\cite{Irkhin:1989}). To this end, we introduce the normalized
spectral spin function
\[
K_{\mathbf{q}}(\omega )=-\frac 1\pi N_{\mathrm{B}}(\omega )\frac{%
\mathop{\mathrm{Im}}\langle \!\langle S_{\mathbf{q}}^z|S_{-\mathbf{q}%
}^z\rangle \!\rangle _\omega }{\langle S_{-\mathbf{q}}^zS_{\mathbf{q}%
}^z\rangle }
\]
where $N_{\mathrm{B}}(\omega )$ is the Bose function. In the far
paramagnetic region (where spin correlations can be neglected) we can use
the simplest spin-diffusion approximation with
\begin{equation}
K_{\mathbf{q}}(\omega )=\frac 1\pi \frac{Dq^2}{\omega ^2+(Dq^2)^2}
\label{eq:K}
\end{equation}
where $D$ is the spin diffusion constant (in fact, the results depend weakly
on the concrete form of spin dynamics). Then we have to put in (\ref{eq:EGF})
\begin{eqnarray}
a_{\mathbf{k}\alpha }(E) &=&P_\alpha +\sum_{\mathbf{q}}\int d\omega K_{%
\mathbf{q}}(\omega )\frac{t_{\mathbf{k-q}}}{(2S+1)^2}\frac{\chi +(2S+1)n_{%
\mathbf{k-q}\alpha }}{E-P_\alpha t_{\mathbf{k-q}}-\omega },  \label{eq:a:K}
\\
b_{\mathbf{k}\alpha }(E) &=&E-\sum_{\mathbf{q}}\int d\omega K_{\mathbf{q}%
}(\omega )\frac{t_{\mathbf{k-q}}^2}{2S+1}\frac{n_{\mathbf{k-q}\alpha }}{%
E-P_\alpha t_{\mathbf{k-q}}-\omega }.  \label{eq:b:K}
\end{eqnarray}
To simplify numerical calculations, we average the spectral function
(\ref{eq:K}) in~$\mathbf{q}$,
\[
K_{\mathbf{q}}(\omega )\rightarrow \overline{K}(\omega )=\sum_{\mathbf{q}}K_{%
\mathbf{q}}(\omega ),
\]
which is sufficient to obtain qualitatively valid results. Indeed, this
approximation (which is in spirit of the large-$d$ or large-$z$ expansion)
reproduces correctly the low-frequency behavior of spin fluctuations which
is important near the Fermi level. Then $a(E)$ and $b(E)$ do not depend
on~$\mathbf{k}$. We have taken below $D=0.7c|t|$ ($t$ is the transfer integral).

In the far paramagnetic region we have
\begin{eqnarray*}
\langle (S_{\mathrm{tot}}^z)^2\rangle &=&\frac 13\langle \mathbf{S}_{\mathrm{%
tot}}^2\rangle =\frac 13\chi =\frac{1-n}{2S+1}\left( \frac 23S^3+S^2+\frac
13S\right) \\
&&+\frac n{2\widetilde{S}+1}\left( \frac 23\widetilde{S}^3+\widetilde{S}%
^2+\frac 13\widetilde{S}\right) .
\end{eqnarray*}
Then we obtain for $S=1/2$
\[
\chi =\left\{
\begin{array}{ll}
3(1-n)/4+2n, & \quad \alpha =+, \\
3(1-n)/4, & \quad \alpha =-.
\end{array}
\right.
\]

\begin{figure}[htbr]
\begin{center}
\epsfig{file= 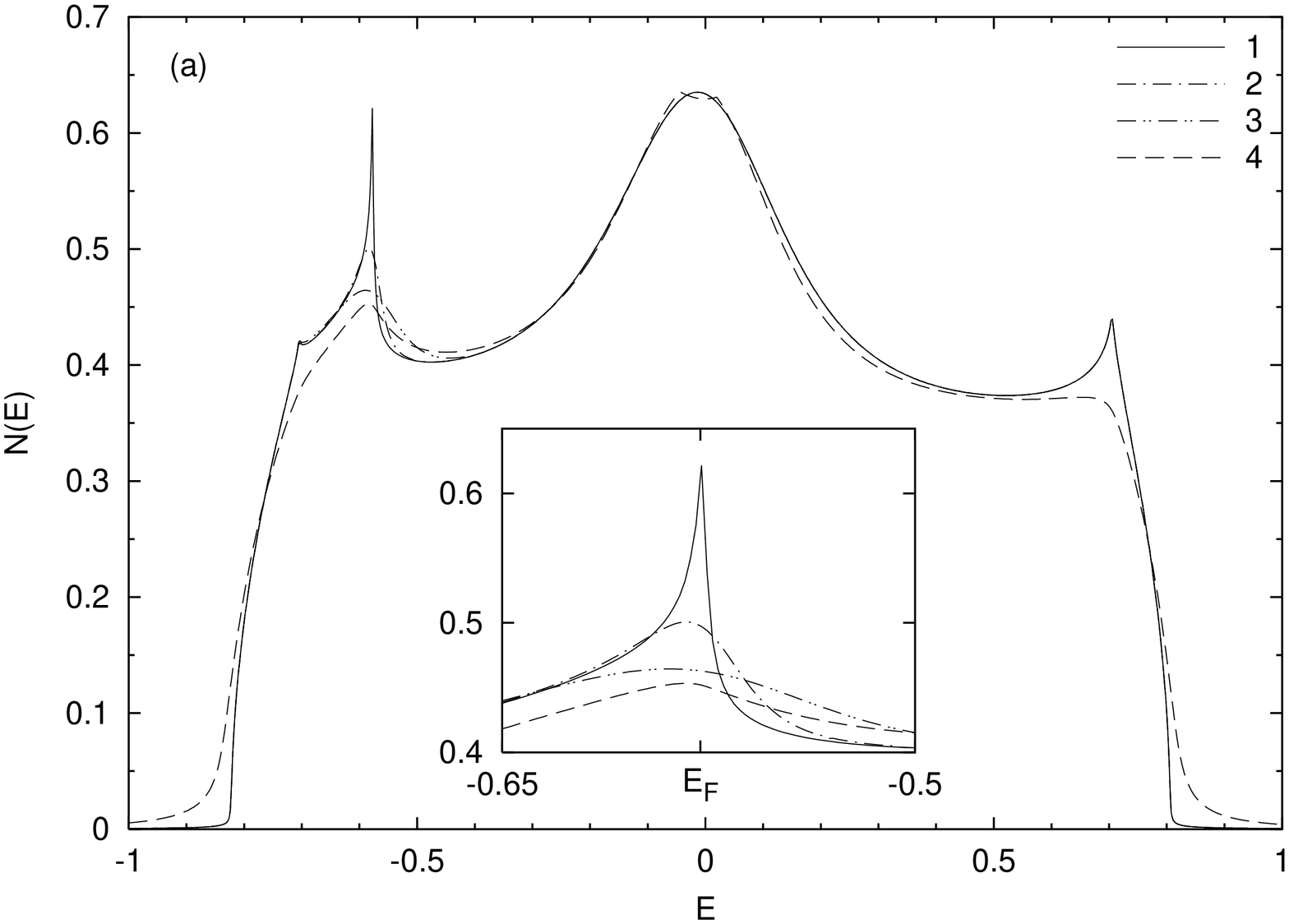, width=0.55\textwidth, angle=-90}
\epsfig{file= 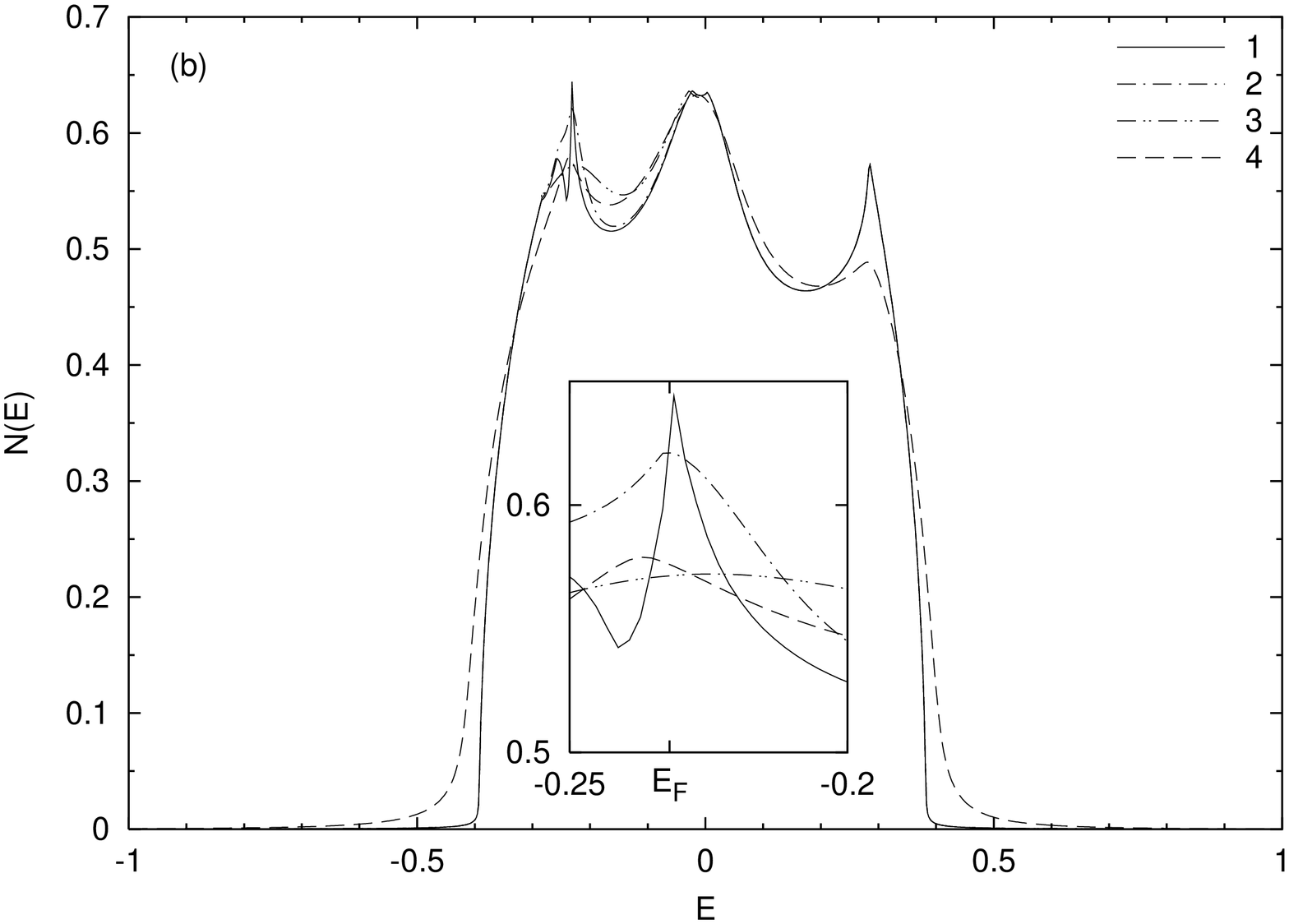, width=0.55\textwidth, angle=-90}
\end{center}
\caption{The density of states picture for the Green's function
(\ref{eq:EGF}) and semielliptic bare DOS with $S=1/2$, $n=0.15$ in the
absence of spin dynamics for different temperatures (Eqs.~(\ref{eq:a}),
(\ref{eq:b}), lines~1, 2, 3 correspond to $T=0$, $0.01$, $0.03$) and with
account of spin dynamics at zero temperature (Eqs.~(\ref{eq:a:K}),
(\ref{eq:b:K}), line~4); (a) $\alpha =+$ and (b) $\alpha =-$. Inset shows the
region near the Fermi energy which corresponds to the peak top. Energy and
temperature is measured in units of bare half-bandwidth.}
\label{fig:1}
\end{figure}

The results of numerical calculation of the single-particle density of
states
\begin{equation}
N_\alpha (E)=-\frac 1\pi \mathop{\mathrm{Im}}\sum_{\mathbf{k}}G_{\mathbf{k}%
\alpha }(E)  \label{eq:DOS}
\end{equation}
for the semielliptic bare DOS are shown in Fig.~\ref{fig:1}. One can see
that pronounced density of states peaks occur at the Fermi level. The peaks
are smeared by both including spin dynamics and finite temperatures. It
should be noted that, unlike the large-$d$ approach \cite{Georges:1996}, the
limit of zero temperature makes no difficulties.

As well as for small-$I$ Kondo problem, the ``Kondo effect'' is connected
with the Fermi functions (which were not treated in Ref.~\cite{Anokhin:1991})
and is a quantum effect that is small in $1/S$. Unlike the FM case, where
the singularity has one-sided form and non-quasiparticle (incoherent)
contributions to the density of state plays main role \cite{Zarubin:1999},
the logarithmic contribution is symmetric with respect to the Fermi level.
Note that after formal expansion in $1/z$ the contributions to the Green's
function with the Fermi functions are canceled in the paramagnetic phase to
first order \cite{Irkhin:1990}.

As discussed in Ref.~\cite{Anokhin:1991}, the approximation (\ref{eq:a}),
(\ref{eq:b}) leads to some formal difficulties connected with occurrence of
an additional singularity of the Green's function in the complex plane. This
can result in the violation of analytical properties, in particular, of the
normalization condition
\begin{equation}
L=\langle \{g_{i\sigma \alpha },g_{i\sigma \alpha }^{\dagger }\}\rangle
=P_\alpha  \label{eq:L}
\end{equation}
where
\[
L\equiv \int\limits_{-\infty }^{+\infty }dEN_\alpha (E).
\]
The analytical properties turn out to be different for the cases of positive
and negative $I$. For $I>0$ the singularity lies in the upper half-plane,
and for $I<0$ in the lower one, so that the normalization condition is
violated for $I<0,$ although this violation is numerically not too large
(see Table~\ref{tab:1}). This difficulty does not seem to lead to unphysical
conclusions in our simple approximation, since the results for the DOS
picture in both the cases are qualitatively similar. However, it should be
stressed that this problem has a rather general character and is typical for
most calculations which use the many-electron Hubbard representation (or
related slave-boson and slave-fermion representations which include
constraint conditions). It is connected with the $\mathbf{k}$-dependence of
``perturbation'' (band energy $t_{\mathbf{k}}$), which changes its sign in
the band, and a complicated structure of the Green's fucntion (the simple
Dyson equation is not valid). By these reasons, the signs of the
corresponding imaginary parts are not fixed; the problems increase when
using self-consistent procedures. In particular, such a difficulty should
occur in the non-crossing approximation (NCA) for the Anderson model where
an expansion in the hybridization $V_{\mathbf{k}}$ is constructed.
Unfortunately, this problem is usually misregarded since the normalization
condition is practically never verified.

\begin{figure}[htbr]
\begin{center}
\epsfig{file= 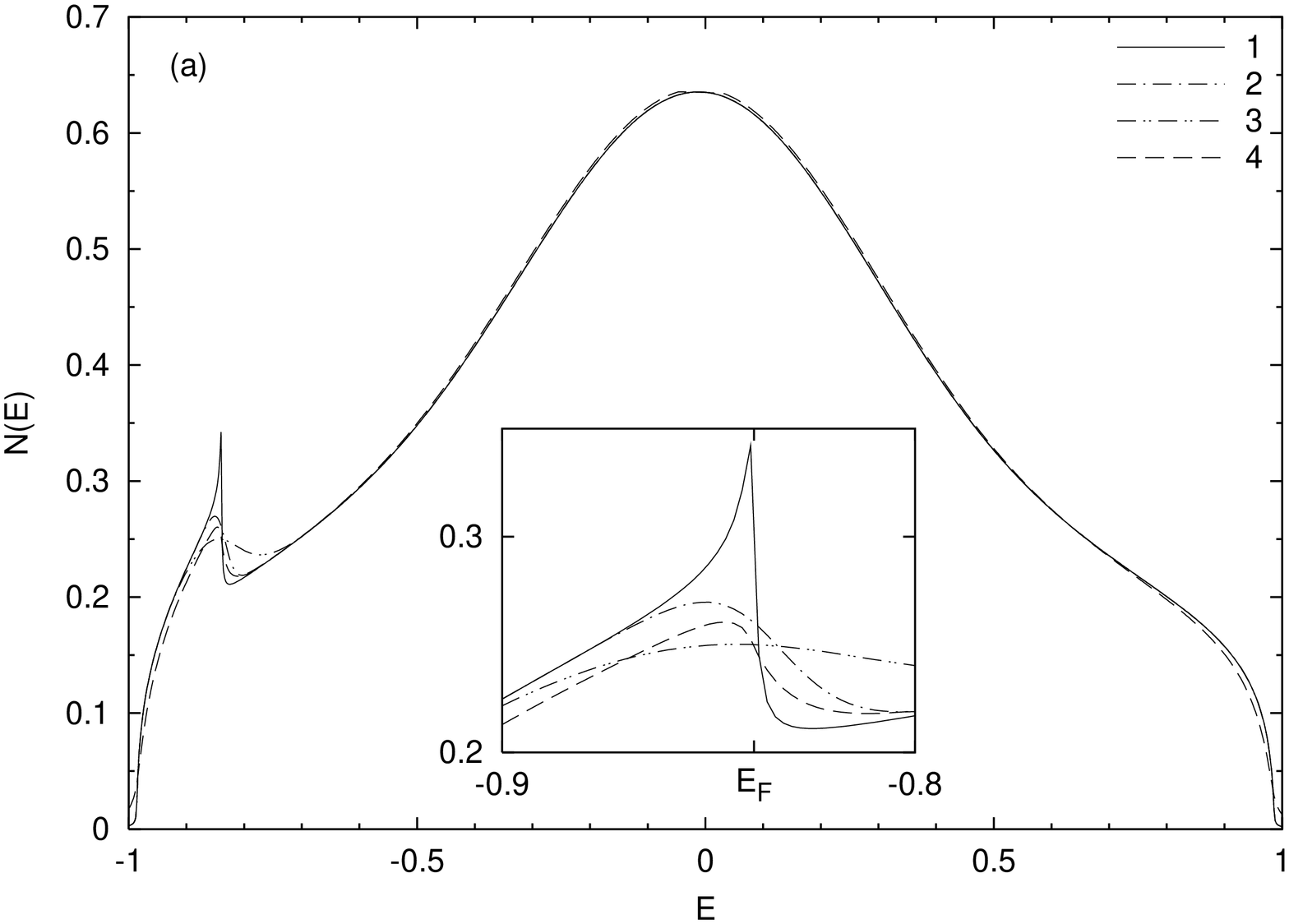, width=0.55\textwidth, angle=-90}
\epsfig{file= 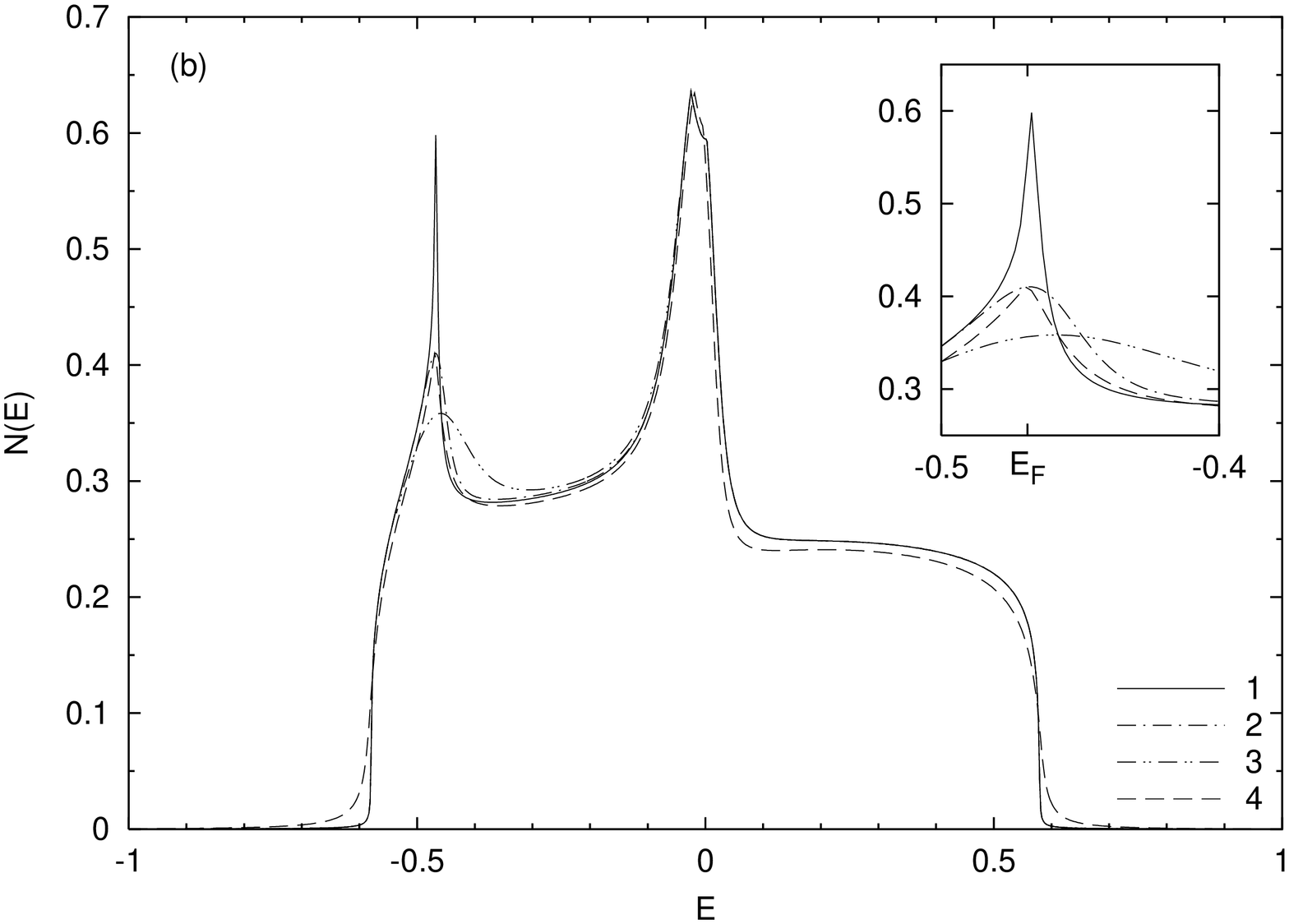, width=0.55\textwidth, angle=-90}
\end{center}
\caption{The density of states picture in the self-consistent approximation
(\ref{eq:s-c1}) with $S=1/2$, $n=0.05$. The notations are the same as in
Fig.~\ref{fig:1}.}
\label{fig:2}
\end{figure}

The approximation (\ref{eq:EGF}), (\ref{eq:a}), (\ref{eq:b}) has an
unpleasant physical drawback: DOS has the ``Van Hove'' singularities at the
energies, corresponding to the edges of the Hubbard-I band
\cite{Anokhin:1991}. To remove this drawback, we renormalize self-consistently
the bandwidths in the resolvents (i.e., in the denominators in (\ref{eq:a:K}),
(\ref{eq:b:K})) and in the Fermi functions by replacing
\begin{equation}
E-P_\alpha t_{\mathbf{k-q}}\rightarrow E-\widetilde{P}_\alpha t_{\mathbf{k-q}%
}.  \label{eq:s-c1}
\end{equation}
Due to this self-consistency procedure, the band edges in the resolvents
coincide with those for the total Green's function (\ref{eq:EGF}). The
corresponding numerical results are shown in Fig.~\ref{fig:2}.The analytical
properties for $I>0$ are not violated (Table \ref{tab:1}). This
approximation is in spirit of large-$N$ expansion and retains the
quasiparticle picture, unlike the Hubbard-III aproximation and the
self-consistent approximation considered below which yield a strongly
incoherent behavior.

\begin{figure}[htbr]
\begin{center}
\epsfig{file= 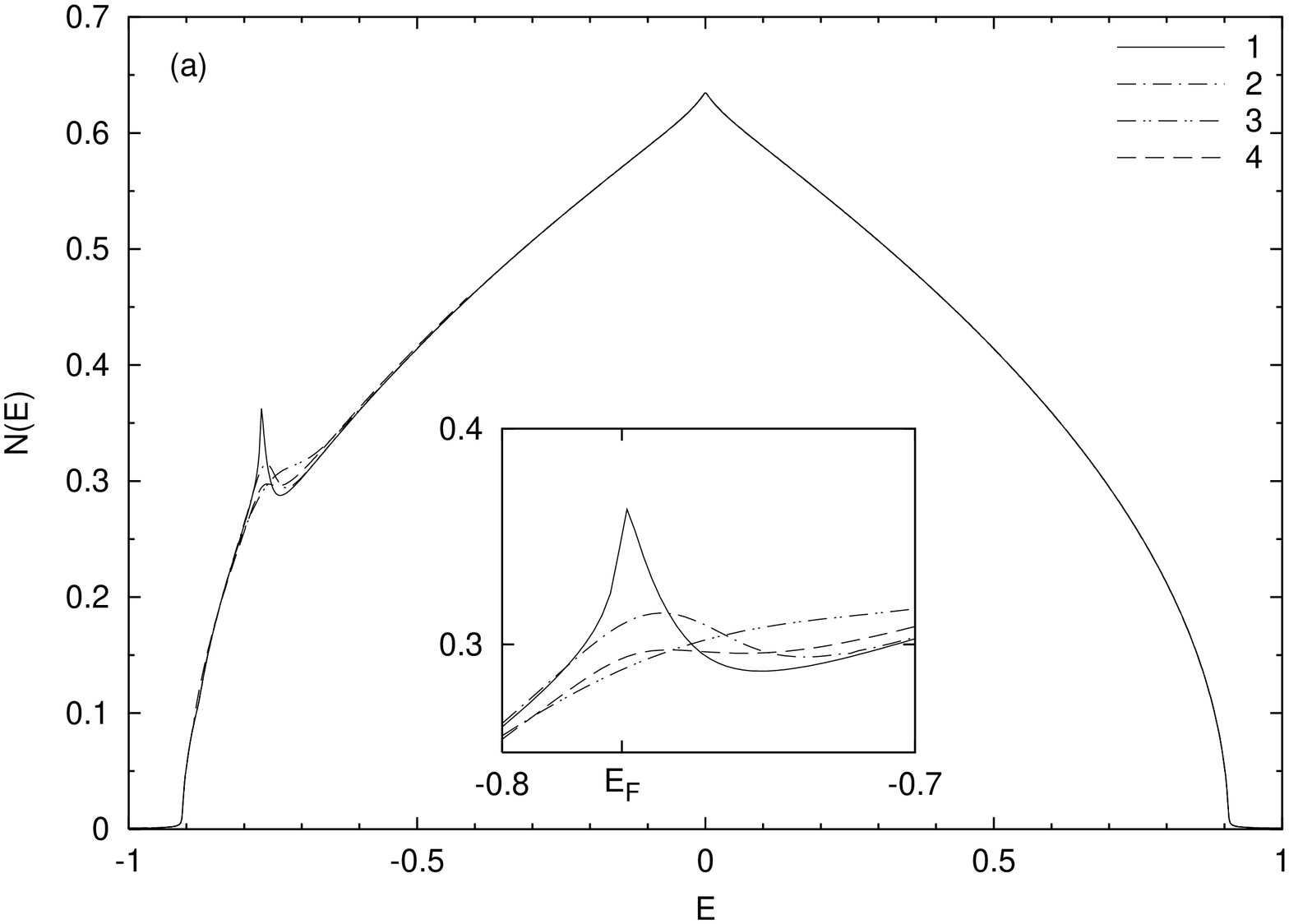, width=0.55\textwidth, angle=-90}
\epsfig{file= 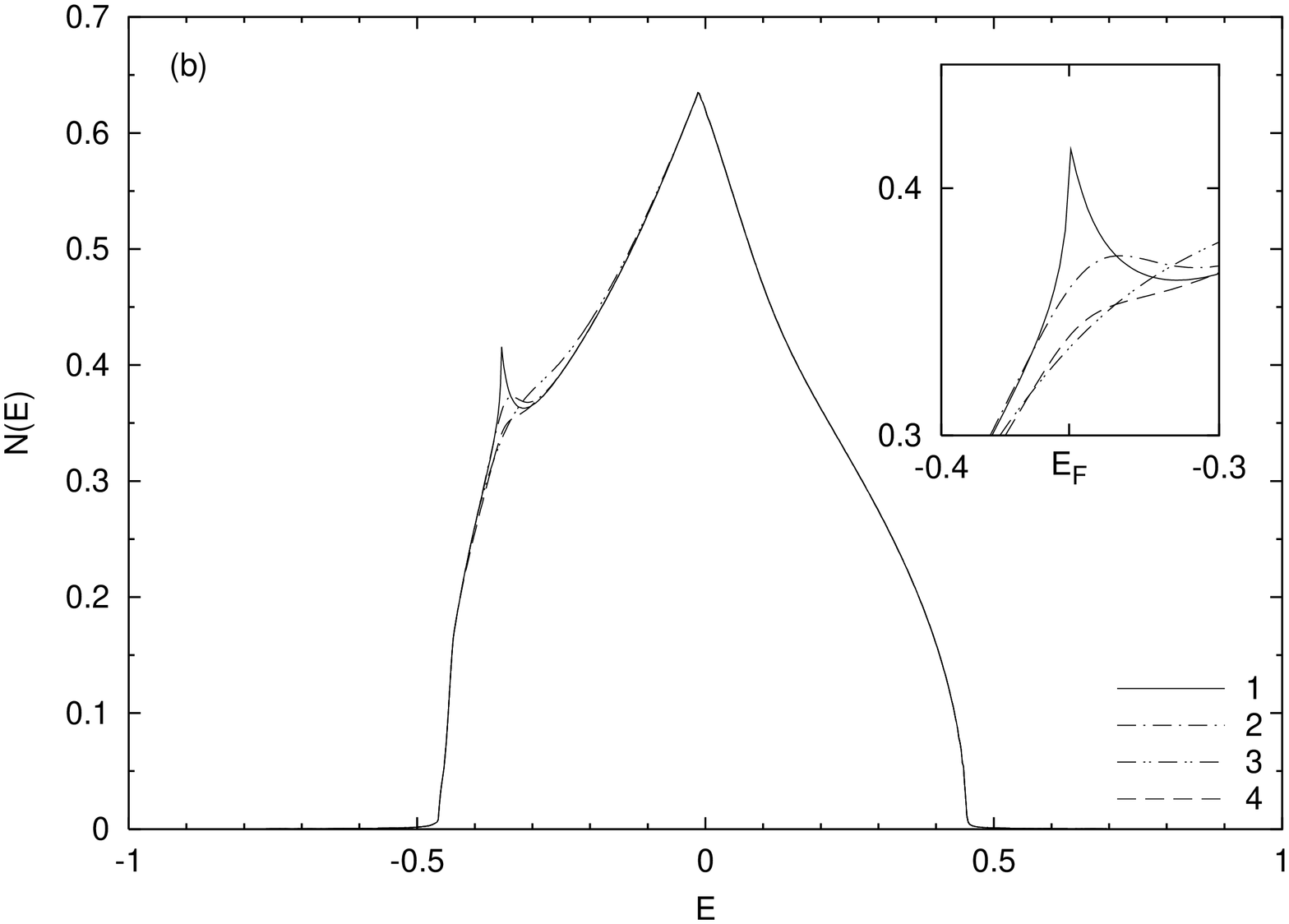, width=0.55\textwidth, angle=-90}
\end{center}
\caption{The density of states picture in the self-consistent approximation
(\ref{eq:s-c}) for $S=1/2$, $n=0.05$.}
\label{fig:3}
\end{figure}

Finally, we discuss also the ``true'' self-consistent approximation where
the exact Green's functions are included into resolvents. Then we have to
replace in the denominators of (\ref{eq:a:K}), (\ref{eq:b:K})
\begin{equation}
E-P_\alpha t_{\mathbf{k-q}}\rightarrow b_{\mathbf{k-q}\alpha }(E)-a_{\mathbf{%
k-q}\alpha }(E)t_{\mathbf{k-q}}  \label{eq:s-c}
\end{equation}
and to make the corresponding renormalization of the distribution functions
\[
n_{\mathbf{k}\alpha }=-\frac 1\pi \int dEf(E)\mathop{\mathrm{Im}}G_{\mathbf{k%
}\alpha }(E).
\]
One can see that the DOS picture (Fig.~\ref{fig:3}) becomes somewhat smeared
even in the absence of spin dynamics, but the smearing is not so strong as
in the FM case \cite{Zarubin:1999}. In this self-consistent approximation,
the analytical properties are violated for both $I>0$ and $I<0$
(Table~\ref{tab:1}).

To conclude, we have analyzed formation of the ``Kondo'' peak in narrow-band
systems at the Fermi level within a simple analytical approach. The
numerical results demonstrate sharp energy dependence of the density of
states near $E_{\mathrm{F}}$. As well as in the standard Kondo problem, one
can expect strong temperature dependences of thermodynamic and transport
properties (e.g., an enhancement of the effective electron mass). This
problem needs further investigations. In particular, summation of
higher-order perturbation corrections with the use of slave boson and
fermion representations would be of interest.

We are grateful to A.O.Anokhin for useful discussions. The research
described was supported in part by the Grant 99-02-16279 from the Russian
Basic Research Foundation.

\eject

\section*{Figure and table captions}

\begin{table}[htbr]
\caption{Values of the normalization factors $L/P_\alpha $ (see Eq. (\ref
{eq:L})) for semielliptic bare DOS in (I) the approximation (\ref{eq:a}),
(\ref{eq:b}); (II) the self-consistent approximation (\ref{eq:s-c1}); (III)
the self-consistent approximation (\ref{eq:s-c}).}
\label{tab:1}
\begin{center}
\begin{tabular}{lllll}
\hline\noalign{\smallskip}
$\alpha $ & $n$ & \multicolumn{3}{c}{$L/P_\alpha $} \\ \noalign{\smallskip}\cline{3-5}\noalign{\smallskip}
& & I & II & III \\ \noalign{\smallskip}\hline\noalign{\smallskip}
$+$ & 0.20 & 1.000 & 1.000 & 1.022 \\
$+$ & 0.15 & 1.000 & 1.000 & 1.017 \\
$+$ & 0.10 & 1.000 & 1.000 & 1.011 \\
$+$ & 0.05 & 1.000 & 1.000 & 1.004 \\
$+$ & 0.02 & 1.000 & 1.000 & 1.002 \\
$+$ & 0.00 & 1.000 & 1.000 & 1.001 \\
$-$ & 0.00 & 1.510 & 1.395 & 1.325 \\
$-$ & 0.02 & 1.489 & 1.366 & 1.328 \\
$-$ & 0.05 & 1.448 & 1.337 & 1.318 \\
$-$ & 0.10 & 1.385 & 1.331 & 1.277 \\
$-$ & 0.15 & 1.342 & 1.246 & 1.227 \\
$-$ & 0.20 & 1.289 & 1.192 & 1.177 \\ \noalign{\smallskip}\hline
\end{tabular}
\end{center}
\end{table}

\end{document}